%% file: final_version.tex
\documentclass[acmsmall]{acmart}

\AtBeginDocument{%
  \providecommand\BibTeX{{%
    \normalfont B\kern-0.5em{\scshape i\kern-0.25em b}\kern-0.8em\TeX}}}

\usepackage{algorithm,algpseudocode}
\usepackage{algorithm,algpseudocode}

\usepackage{soul}
\received{February 2019} 
\received[revised]{April 2019}
\received[accepted]{May 2019}

\begin{document}

\setcopyright{acmlicensed}
\acmJournal{POMACS}
\acmYear{2019}\acmVolume{3}\acmNumber{2}\acmArticle{41}\acmMonth{6}
\acmPrice{15.00}
\acmDOI{10.1145/3326156}

\title{QuickStop: A Markov Optimal Stopping Approach for Quickest Misinformation Detection}

\author{Honghao Wei}
\affiliation{%
  \institution{Arizona State University}
  \city{Tempe}
  \state{Arizona}
  \postcode{85287}
}
\email{hwei30@asu.edu}

\author{Xiaohan Kang}
\affiliation{%
  \institution{University of Illinois at Urbana-Champaign}
  \city{Urbana}
  \state{Illinois}
  \postcode{61820}
}
\email{Veggente@gmail.com}

\author{Weina Wang}
\affiliation{%
  \institution{Carnegie Mellon University}
  \city{Pittsburgh}
  \state{Pennsylvania}
  \postcode{15213}
}
\email{weinaw@cs.cmu.edu}

\author{Lei Ying}
\affiliation{%
 \institution{Arizona State University}
  \city{Tempe}
  \state{Arizona}
  \postcode{85287}
}
\email{lei.ying.2@asu.edu}
\renewcommand{\shortauthors}{H.\ Wei et al.}

\begin{abstract}
This paper combines data-driven and model-driven methods for real-time misinformation detection. Our algorithm, named \textsc{QuickStop}, is an optimal stopping algorithm based on a probabilistic information spreading model obtained from labeled data. 
The algorithm consists of an offline machine learning algorithm for learning the probabilistic information spreading model and an online optimal stopping algorithm to detect misinformation. The online detection algorithm has both low computational and memory complexities.   Our numerical evaluations with a real-world dataset show that \textsc{QuickStop} outperforms existing misinformation detection algorithms in terms of both accuracy and detection time (number of observations needed for detection). Our evaluations with synthetic data further show that \textsc{QuickStop} is robust to (offline) learning errors.
\end{abstract}



\maketitle

\input{intro}

\input{model}
\input{analysis}

\input{alg-eval}

\input{related}
\input{conclusions}
\section*{Acknowledgement}
The authors thank R.\ Srikant for his invaluable comments and feedback. This work was supported in part by NSF Grant IIS-1715385.

\bibliographystyle{ACM-Reference-Format}
\bibliography{bibliography1,inlab-refs}

\input{app}
\end{document}

%% file: intro.tex
\section{Introduction}
The proliferation of misinformation \cite{lazbauben_18} (colloquially known as ``fake news'') on online social networks has become one of the greatest threats to our national security, has eroded the public trust in news media, and is an imminent threat to the ecosystem of online social platforms like Facebook, Twitter and Sina Weibo. For example, in 2013, a fake tweet claiming that the then US President Barack Obama was injured by explosives from a hacked Twitter account of the Associated Press caused a 150-point drop of the Dow Jones in just two minutes;\footnote{\url{https://www.washingtonpost.com/news/worldviews/wp/2013/04/23/that-tipped-stock-market-by-136-billion-is-it-terrorism/}} and fake news in the 2016 US Presidential Election has led to increased political and social polarization and posed a great threat to democracy.  Social media companies, such as Facebook and Twitter, are now taking multiple countermeasures to combat misinformation as the proliferation of misinformation is driving users away from these platforms.

Despite the enormous attention it receives and the tremendous efforts from both public and private institutions to counter it,  misinformation detection remains a daunting task as of today.   Online platforms and news organizations have experimented different methods. Facebook launched its fact-checking project in Spring 2018 to work with third-party publishers to validate facts and accuracy of news articles.\footnote{\url{https://www.facebook.com/help/1952307158131536?helpref=faq_content}} The New York Times has recently published a tip form so that its readers can report misinformation and fake news.\footnote{\url{https://www.nytimes.com/2018/09/17/technology/disinformation-tipsheet.html}} The third-party fact-checking method is often very effective for detecting whether a specific news article is fake or not, but clearly is not a scalable solution and cannot cover even a tiny fraction of news articles and tweets (there are about 500 million tweets per day on Twitter). The crowdsourcing approach used by New York Times is more scalable, but the reports are not always trustworthy because anyone can send a tip. In light of these challenges, machine-learning and data-mining approaches have emerged to tackle misinformation detection in a systematic way (see \cite{shusliwan_17} for a comprehensive review). It has been shown in \cite{casmenpob_11} that the features extracted from the content of a news article, the features of the users who spread the news, and the connections of these users can be effectively utilized for misinformation detection. These are exciting discoveries and progresses because ``machine-based'' methods are much more scalable than ``human-based'' methods, and can handle a vast number of news articles in a short period of time.

While machine-learning approaches address the scalability issue, another important aspect of misinformation detection, \emph{speed} or \emph{sample complexity} (the amount of time or the number of observations needed to detect misinformation), has yet to be tackled.  Speed is important because of the disruptive nature of misinformation, which often causes significant damages in a very short period of time. For example, it only took less than two minutes to tip the Dow Jones by 150 points with one single fake tweet. Therefore, it is imperative to detect misinformation at the earliest time so that proper countermeasures can be taken to suppress it. A fact-checking approach may take a few hours because fact-checkers need to gather facts and evidence to validate or invalidate a news article. Therefore, the speed aspect of misinformation detection is equally important as accuracy and scalability in the design of misinformation detection algorithms.

Motivated by the discussions above, this paper focuses on quickest detection of misinformation. The goal is to develop an algorithm that addresses the three important considerations in misinformation detection: scalability, accuracy and {\em speed}. Note that existing machine-learning-based approaches have demonstrated a strong correlation between user features and the spreading models under different information types (real or fake). We will demonstrate this strong correlation in Section~\ref{sec:model} using a Sina Weibo dataset. The signal of a single retweet is often very weak and usually not sufficient for classifying a news article with a reasonable accuracy. But this accuracy can be improved with more and more weak signals. This paper views the problem of misinformation detection as a \emph{sequential} hypothesis testing problem.  As the platform receives a \emph{sequence} of weak signals in real time, it determines whether it has collected enough information to declare the type of the news (real or fake). The more signals collected, the more accurate the detection result will be, but then we are at risk of letting the misinformation spread. Enlightened by these observations, we propose \textsc{QuickStop}, a scalable algorithm that performs accurate, quick detection of misinformation. \textsc{QuickStop} combines a data-driven approach with a model-driven approach in the following way.

\begin{itemize}
\item {\bf Data-based probabilistic modeling}: Since each retweet is a weak signal for the hypothesis testing (whether the news article is real or fake), extracting the statistics of these weak signals is important for establishing an effective probabilistic model for hypothesis testing.  \textsc{QuickStop} first uses an SVM (Support Vector Machine) algorithm to extract an edge-based probabilistic information spreading model. Section~\ref{sec:model} explains the rationale behind the edge-based model (compared with a node-based model) and shows the effectiveness using the Sina Weibo dataset.

\item {\bf Model-based quickest detection}: After establishing the probabilistic model, we formulate the quickest misinformation detection problem as an \emph{optimal stopping problem}. Specifically, we propose a cost model that includes both the cost due to detection error and the cost due to the propagation of misinformation.  Note that the propagation cost occurs only in the case of misinformation.  With this formulation, the goal is to discover a \emph{stopping policy}, i.e., a policy that determines when to stop collecting observations and what type to declare after stopping, that minimizes the overall cost. As more observations are collected, the error cost decreases but the propagation cost could increase in the case of misinformation.  Therefore, the optimal stopping policy needs to balance the detection accuracy and detection time so that misinformation can be detected confidently at the earliest possible time.
\end{itemize}
The main contributions of this paper are summarized below.
\begin{itemize}
\item {\bf Problem Formulation}: We formulate the quickest misinformation detection problem as a Markov optimal stopping problem based on a probabilistic information spreading model.  This probabilistic model can be extracted from training datasets by given classifiers.  An interesting feature of our formulation is the asymmetric cost functions between real news and misinformation --- spreading misinformation causes far more damage than spreading real news so we need to act quickly only in the case of misinformation.  The analytical solution of a Markov optimal stopping problem in general requires computing a function of the state (see, e.g., Chapter 3.4.4 on Page~59 of \cite{poovinhad_09}).  Since the state in our formulation includes the current time index (i.e., how many users the article has reached), this function would be \emph{time-dependent}.  Effectively, this means that we potentially need a different function of the collected information for each time step.  However, utilizing structures in our probabilistic model, we show that the optimal stopping policy has a simple threshold form described by several \emph{time-independent} thresholds.  We comment that this structure is similar to that in the solution of the sequential testing problem, but the techniques there do not directly apply to our problem since our cost function has a nonlinear term due to the asymmetry.

\item {\bf Algorithm and Analysis}: We propose an algorithm named \textsc{QuickStop} that detects misinformation based on \emph{edge} types, where an edge is a connection between two individuals along which a piece of information spreads from one individual to the other. \textsc{QuickStop} consists of two parts: (i) \textsc{QuickStop}-Training, an offline algorithm that classifies edges into four types and then calculates transition probabilities between different edge types, where the transition probabilities in the case of real news may be different from those for misinformation; and (ii) \textsc{QuickStop}-Detection, an online detection algorithm with low computational and memory complexities. We emphasize that the main computation load is in the offline part.  Once the offline training is completed, the online part for detection is very efficient as described below.  \textsc{QuickStop}-Detection maintains a scalar variable that describes the current state, and updates the state for each new observation. The update just follows a simple formula and its complexity does not depend on how many observations have been collected. Then the algorithm compares the state with several thresholds calculated offline.  Based on the comparison result, it decides whether it will keep collecting observations or declare the type of the information.  In the latter case, what type to declare is also determined by the comparison result.  Therefore, \textsc{QuickStop}-Detection has very low computational and memory complexities, and is ideal for real-time large-scale misinformation detection.

\item {\bf Evaluations}: We evaluated the performance of \textsc{QuickStop} using both a real-world social network dataset (from Sina Weibo) and synthetic data. The evaluations on the real-world dataset demonstrates the effectiveness of our algorithm in terms of both accuracy and speed compared with state-of-the-art real-time misinformation algorithms. Under \textsc{QuickStop} with a low propagation cost, it took  12 observations on average in the Weibo dataset to detect misinformation, but more to declare real news. This is consistent with the asymmetric cost model. Furthermore, the false negative rate (misinformation classified as real news) is much lower than the false positive rate (real news classified as misinformation), which is also desirable in practice. In contrast, the accuracy of the state-of-the-art early detection algorithms are still lower than ours even with $33\times$ more observations. 
    From the evaluations on synthetic data, we further observed that \textsc{QuickStop} is robust to classification errors.
    \end{itemize}

We finally comment that while several early misinformation detection algorithms have been developed   \cite{zharesmei_15,magaozho_15,magaopra_16,magaowon_17,magaokam_18,chexuehon_18}, these algorithms either use a fixed number of observations as input \cite{magaokam_18,chexuehon_18} or observations over a fixed time period as input \cite{magaozho_15,magaopra_16,zharesmei_15,magaowon_17}. Therefore, these early detection algorithms  do  not minimize the detection time (or the number of observations) in real time. Our approach, on the other hand, tackles the problem using the optimal stopping method and optimizes the number of observations needed in real-time for quickest detection.  Our numerical evaluations show \textsc{QuickStop} achieves higher accuracy with fewer observations due to the dynamic nature of the algorithm. A detailed review of other related work is presented in Section \ref{sec:related}.

%% file: model.tex
\section{Model and Problem Statement}
\label{sec:model}
We model an online social network as a graph $G=(\mathcal{V},\mathcal{E}),$ where $\mathcal{V}$ is the set of vertices representing users and $\mathcal{E}$ is the set of directed edges representing the connections between users. Information (real news or misinformation) can spread from one user to another via the edge connecting them, e.g., a Twitter user can retweet a post from one of her/his followees. In this paper, we adopt the terminology of Twitter. Given a directed edge $(v,u),$ user $u$ is called a follower of user $v;$ and user $v$ is called a followee of user $u.$  Information can spread from user $v$ to user $u$ via this directed edge.

We assume two types of information that may spread in the network: {\em real news articles} (simply called {\em news} in the remainder of the paper) and {\em misinformation}. A user (say user $u$) decides whether to post (retweet) the information based on the following three factors: (i) the type of the information, (ii) the features of user $u$, and (iii)  the set of user $u$'s neighbors who have posted (retweeted) the information before user $u.$

As information spreads in the network, the platform obtains sequential observations (weak signals) for misinformation detection. In this paper, a retweet is considered to be an observation, which is represented by the edge over which this retweet occurs. Specifically, we define the $k$th observation to be  $({\bf V}_k, {\bf U}_k),$ where ${\bf U}_k$ is the feature vector of the $k$th user who retweets the information and ${\bf V}_k$ is the feature vector of the followee from whom the $k$th user retweets the information.  We remark that when complete network and information diffusion information is known, the information spreading trace is likely to be a tree or a forest (with multiple information sources). However, in practice, it is often not the case because of missing information and partial observations \cite{kwochajun_13,jindousar_13}. Therefore, the observations we have are a sequence of retweets $({\bf V}_k, {\bf U}_k),$ which not necessarily form a tree. In particular, ${\bf U}_k$ is not necessarily the same as ${\bf V}_{k+1}$ in the trace. Now to model these retweets as weak signals, we can consider the following two approaches.
\begin{itemize}
\item User-based Model: In the user-based model, given the type of an article, the probability a user retweets the article depends on the features of the user. Intuitively, an honest user has a lower probability to retweet some misinformation than a malicious user (e.g., a bot). The user-based model is to classify the users based on the user features with a labelled training dataset.

\item Edge-based Model: In the edge-based model, we view each edge as a communication channel and classify edges into different groups. For example, misinformation is more likely to spread over an edge between two malicious socialbots than an edge between two honest users. The edge-based model is to classify the edges based on the edge features (the feature vectors of the two end users $({\bf V}, {\bf U})$) with a labelled training dataset.
\end{itemize}

\begin{figure}[h]
	\centering
	\includegraphics[scale=0.5]{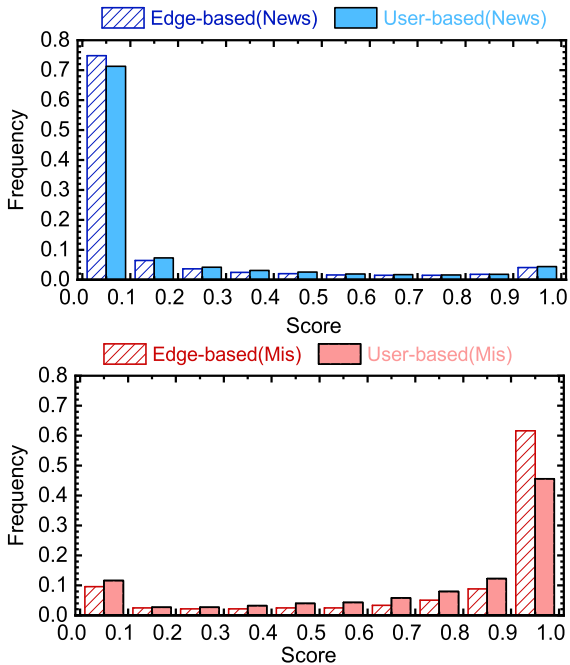}
	\caption{Classification distribution}\label{fig:clas_dis}
\end{figure}

Figure~\ref{fig:clas_dis} presents the distributions of SVM classification scores of the user-based model and the edge-based model of the Weibo dataset released in \cite{magaozho_15}, where $x$-axis is the classification score of the SVM classifier, and $y$-axis is the score distributions (frequencies). A user or an edge with a higher score is considered more likely to spread misinformation. From the figure, we first observe that the scores of users (or edges) involved in spreading news concentrate around zero while the scores of users (or edges) involved in spreading misinformation concentrate around one. This demonstrates a strong correlation between article types and user/edge features. Furthermore, we can see that the score distributions based on edges exhibit a stronger correlation with article types than the score distributions based on users. For example, for misinformation, the score distribution based on edges has a higher frequency around zero than that based on users (60\% versus 45\%). Because of this observation, in this paper, we use the edge-based model.

\begin{figure}[h]
\centering
\includegraphics[scale = 0.8]{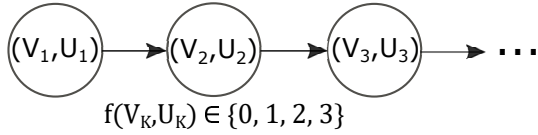}
\caption{A Markov chain model for sequential observations}\label{fig:seq-obs-mc}
\end{figure}

We assume that given the article type, the sequential observations form a Markov chain as shown in Figure~\ref{fig:seq-obs-mc}, where we further assume the edge feature vector $({\bf V}, {\bf U}) $  can be classified into four classes $Z_k=f({\bf V}_k, {\bf U}_k) \in \{0, 1, 2, 3\}$ to simplify the model, where 0 is the type of edges that are most likely to be used for spreading news and 3 is the type of edges that are most likely to be used for spreading misinformation.
Under this Markov chain model, besides the edge types, additional parameters to be learned are the transition probabilities, denoted by $\alpha_i(Z_k|Z_{k-1}),$ where $i\in\{0,1\},$ $i=0$ indicates these are the transition probabilities when spreading news, and $i=1$ indicates these are the transition probabilities when spreading misinformation. When $\alpha_0(\cdot|\cdot)$ and $\alpha_1(\cdot|\cdot)$ are different, we can detect misinformation using sequential hypothesis testing. We remark that the generalization of the four-class model to a $C$-class model for a finite $C$ is straightforward. Our main results and the proposed algorithm work for any finite $C.$ The choice of the number of classes, $C$, however, needs to balance the detection performance, which favors a larger $C,$ and the training complexity and accuracy, which often favor a smaller $C.$ We adopt the four-class model based on experimental evaluations on the Weibo dataset. The evaluations showed that the four-class model performs significantly better than a two-class model, but increasing $C$ from four to eight did not yield any noticeable improvement.

Tables \ref{tab:tr_link-news} and \ref{tab:tr_link-mis} show the empirical transition probability matrices under news and misinformation obtained from the Weibo dataset (an edge with a classification score below $0.25$ is placed in class $0$, one with score between $0.25$ and $0.5$ is placed in class $1$, one with score between $0.5$ and $0.75$ is placed in class $2$, and one with score above $0.75$ is placed in class $3$). We can clearly observe that the observations are not i.i.d., which supports our edge-based Markovian model.
\begin{table}[ht]
	\centering
	\begin{tabular}{|c|c|c|c|c|}
		\hline
		&0&1&2&3\\
		\hline
		0& 0.828& 	0.120& 	0.039& 	0.012\\
		\hline
		1 & 0.651	& 0.224& 	0.084& 	0.041\\
		\hline
		2& 0.500& 	0.193	& 0.191	& 0.116 \\
		\hline
		3&  0.279	& 0.181& 	0.211& 	0.329\\
		\hline
	\end{tabular}
	\caption{Edge Transition Probability Matrix under News from the Weibo Dataset}\label{tab:tr_link-news}
\end{table}
\begin{table}[ht]
	\centering
	\begin{tabular}{|c|c|c|c|c|}
		\hline
		&0&1&2&3\\
		\hline
		0& 0.163& 	0.167& 	0.249& 	0.421 \\
		\hline
		1& 0.105& 	0.194& 	0.239& 	0.461\\
		\hline
		2& 0.080	& 0.119	& 0.277& 	0.524\\
		\hline
		3& 0.052	& 0.088	& 0.203	& 0.657\\
		\hline
	\end{tabular}
	\caption{Edge Transition Probability Matrix under Misinformation from the Weibo Dataset}\label{tab:tr_link-mis}
\end{table}

For the edge classifier,  we leverage the existing research, in particular, the research in \cite{kwochajun_13}, where it shows that SVM performs the best among several popular machine-learning algorithms, including decision tree and random forest for classifying misinformation. We adopt SVM and the user features proposed in \cite{casmenpob_11} to obtain an edge classifier. The details can be found in Section \ref{sec:alg}. After classifying the edges in the training data, we further obtain transition probabilities $\alpha_i(Z_k|Z_{k-1})$ from the training data to build a probabilistic information spreading model (details can be found in Section \ref{sec:analysis}).

Our focus is on the quickest detection formulation after training the edge classifier and learning the transition probabilities $\alpha_i(Z_k|Z_{k-1}).$ In the next section, we will formulate the quickest misinformation detect problem and prove that the problem is a Markov optimal stopping time problem and its solution is a time-invariant threshold policy. Furthermore, the thresholds can be efficiently calculated offline based on the probabilistic model. The online algorithm is of constant computational and memory complexities, and is very easy to implement.

%% file: analysis.tex
\section{Optimal Stopping Approach for Quickest Misinformation Detection}
\label{sec:analysis}
Consider an online social network platform that is monitoring the spread of some information in the network. We say that an event occurs when a user retweets or posts the information. When the $k$th event occurs, we obtain an observation $Z_k\in\{0, 1, 2, 3\}$ by using the trained classifier to learn the edge type. Furthermore, we assume that we have learned the transition probabilities $\alpha_i(Z_k|Z_{k-1})$  from training data.

With the model introduced above, the detection of misinformation can be formulated as a hypothesis testing problem with the following two hypotheses:
\begin{itemize}
\item $H_0$: The information is news. In this case, $\{Z_{k}\}$ is a four-state Markov process with transition probabilities $\alpha_0(Z_k|Z_{k-1}).$

\item $H_1$: The information is misinformation. Then $\{Z_{k}\}$ is a four-state  Markov process with transition probabilities $\alpha_1(Z_k|Z_{k-1}).$
\end{itemize}

Given observations $\{Z_k\},$ the misinformation detection problem is to determine whether $H_0$ or $H_1$ is true. We assume that in terms of the prior distribution, hypothesis $H_0$ occurs with probability $\pi_0$ and $H_1$ occurs with probability $\pi_1=1-\pi_0$. We assume the first observation $Z_1$ is uniformly distributed over $\{0, 1, 2, 3\}$ regardless of the hypothesis.

Now define
\begin{equation}\label{eq:Pik}
    \Pi_{k}=\Pr\left(H_1|(Z_1, Z_2,\cdots, Z_k)\right),
\end{equation}
so $$1-\Pi_{k}=\Pr\left(H_0|(Z_1, Z_2,\cdots, Z_k)\right).$$
According to the Bayes rule, we have
\begin{align*}
\Pi_{k}&=\Pr\left(H_1|(Z_1, Z_2,\cdots, Z_k)\right)\\
&=\frac{\Pr\left((Z_1, Z_2,\cdots, Z_k)|H_1\right) \Pr(H_1)}{\Pr(Z_1,Z_2,\dots,Z_k)}\\
&=\frac{\pi_1\prod_{i=1}^{k-1}\alpha_1(Z_{i+1}\vert Z_i)}{(1-\pi_1)\prod_{i=1}^{k-1}\alpha_0(Z_{i+1}\vert Z_i)+\pi_1\prod_{i=1}^{k-1}\alpha_1(Z_{i+1}\vert Z_i)}.
\end{align*}
From the equation above, we have
\begin{align*}
\frac{1-\Pi_{k}}{\Pi_{k}}&=\frac{(1-\pi_1)\prod_{i=1}^{k-1}\alpha_0(Z_{i+1}\vert Z_i)}{\pi_1\prod_{i=1}^{k-1}\alpha_1(Z_{i+1}\vert Z_i)},
\end{align*} which implies that
\begin{align*}
\frac{1-\Pi_{k+1}}{\Pi_{k+1}}&=\frac{1-\Pi_{k}}{\Pi_{k}}\frac{\alpha_0(Z_{k+1}\vert Z_k)}{\alpha_1(Z_{k+1}\vert Z_k)}.
\end{align*} Therefore, we have the following recursive equation
\begin{eqnarray}
\Pi_{k+1}=\frac{\Pi_{k}\alpha_1(Z_{k+1}\vert Z_k)}{\left(1-\Pi_{k}\right)\alpha_0(Z_{k+1}\vert Z_k)+\Pi_{k}\alpha_1(Z_{k+1}\vert Z_k)}.\label{eq:Pi}
\end{eqnarray} for updating our belief on $H_1.$

Note that given the observation sequence $\{Z_k\},$ we can calculate $\frac{1-\Pi_{k}}{\Pi_{k}}$ in real time. The question is when to declare the type of the information. The more observations we have, the more accurate the decision would be but the more widely the information would have spread. Therefore, we need to balance the accuracy and the potential damage of spreading misinformation. Let $T \ge 1$ denote the random time at which the type of information is declared, which is a function of $Z_1, Z_2, \cdots, Z_T$; i.e., $T$ is a \emph{stopping time} with respect to $\{Z_k\}$.  Let $\delta_T$ denote the type of information that is declared by a detection algorithm. We consider the following two types of costs in the misinformation detection problem.

\subsection*{Error Cost}

The first type of cost comes from mis-detection. Let $c_{\textnormal{I}}$ denote the cost of type-I error (also called false positive, where news is declared as misinformation) and $c_{\textnormal{II}}$ denote the cost of type-II error (also called false negative, where misinformation is declared as news). The expected cost of mis-detection is $$c_e(\delta_T)=c_{\textnormal{I}}\Pr(\delta_T=1|H_0)({1-\pi_1})+ c_{\textnormal{II}}\Pr(\delta_T=0|H_1){\pi_1} ,$$ where $\pi_1$ is the prior probability of $H_1.$

\subsection*{Propagation Cost}
The other type of cost is the propagation cost.  Information becomes more influential when more people share it. So we need to detect misinformation as quickly as we can to limit its potential damage, while spreading news does not occur any cost. Consequently, the propagation cost in our model is asymmetric and comes only from misinformation. In particular, we assume that there is a cost of $c$ associated with each time slot of propagation if the information is misinformation. Thus, at the stopping time $T$, the propagation cost is
$$E\left[cT{\mathbb I}_{H_1}\right],$$
where ${\mathbb I}_{H_1}$ is the indicator function which is equal to $1$ when $H_1$ is true and is equal to $0$ when $H_0$ is true.

\subsection*{A Markov Optimal Stopping Approach}
The goal of the misinformation detection algorithm is to minimize the overall cost. Formally, we aim to find a stopping time $T$ and a decision rule $\delta_T,$ both depending on $Z_1, \cdots, Z_T,$ that solve the following problem
\begin{equation}\label{eq:misinfodetection}
\inf_{T,\delta_T} c_e(\delta_T)+E\left[cT{\mathbb I}_{H_1}\right].
\end{equation}
{An important step for} solving this problem is to properly handle the propagation cost term $E\left[cT{\mathbb I}_{H_1}\right],$ which depends on the hypothesis.  Note that if this term were $E\left[cT\right]$, this problem would be the same as the renowned \emph{sequential testing problem} (see, e.g., \cite{poovinhad_09}).  Specifically, the sequential testing problem solves
\begin{align}
\inf_{T,\delta_T} c_e(\delta_T)+E\left[cT\right].\label{eq:seq}
\end{align}

{Recall that given ${\mathbb I}_{H_1}$, the observation sequence $\{Z_k\}$ is a Markov chain.  Therefore, when we view ${\mathbb I}_{H_1}$ as part of the state, $\{({\mathbb I}_{H_1},Z_k)\}$ forms a Markov chain, and thus the formulation \eqref{eq:misinfodetection} is a Markov optimal stopping problem. However, this Markov chain is only \emph{partially observable} since we cannot observe ${\mathbb I}_{H_1}$.  We can transform this optimal stopping problem of a partially observable Markov chain to a fully observable optimal stopping problem.  Specifically, consider the conditional distribution of ${\mathbb I}_{H_1}$ given observations $Z_1,Z_2,\dots, Z_k$.  This conditional distribution can be represented by the variable $\Pi_k$ defined in \eqref{eq:Pik}.  We can verify that $\{(\Pi_k, Z_k)\}$ is a Markov chain.  For the convenience of analysis, we also view the time index $k$ as part of the state and consider the Markov chain $\{(\Pi_k, Z_k, k)\}$.  With this, we transform the optimal stopping problem in \eqref{eq:misinfodetection} to a Markov optimal stopping problem in Theorem~\ref{THM:1}.}

\begin{theorem}
	The optimal stopping problem \eqref{eq:misinfodetection} is equivalent to a Markov optimal stopping problem with respect to the Markov chain $\{(\Pi_k, Z_k, k)\}$.  Formally,
\begin{align}
&\inf_{T,\delta_T} c_e(\delta_T)+E\left[cT{\mathbb I}_{H_1}\right]\nonumber\\
=&\inf_{T\in\mathcal{T}}E\left[\min\{c_{\textnormal{II}}\Pi_{T},c_{\textnormal{I}}(1-\Pi_{T})\}+cT\Pi_{T}\right],\label{eq:Markov}
\end{align}
where $\mathcal T$ is the set of stopping times with respect to $\{(\Pi_k, Z_k, k)\}$.
\label{THM:1}\hfill{$\square$}
\end{theorem}

Note that the variable in the Markov stopping problem \eqref{eq:Markov} is just the stopping time $T$ instead of both $T$ and $\delta_T.$  Therefore, we can find the optimal stopping policy in two steps: first find the optimal stopping time $T$ by solving \eqref{eq:Markov}, and then find the optimal decision rule $\delta_T$ based on $Z_1, \dots Z_T$. Such a transform from a partially observable Markov chain to a fully observable Markov chain has been widely used in optimal stopping problems and more generally in Markov decision processes (see, e.g., Chapter 4.1 in Vol.\ I of \cite{Ber_17}, and \cite{YeZho_13}).  Here we include the proof of Theorem~\ref{THM:1} in Appendix~\ref{app:proof:thm1} for completeness.

The analytical solution of the Markov optimal stopping problem \eqref{eq:Markov} can be obtained using the Snell envelope (see, e.g., Chapter 2.2 on Page~38 of \cite{LamLap_08}, and Chapter 3.4.4 on Page~59 of \cite{poovinhad_09}), which, in general, needs to compute a function of the state and store the function for use in the optimal stopping policy.  In our problem, the state includes the time index $k$.  Then to compute the Snell envelope, we potentially need a different function of the collected information $\Pi_k$ and $Z_k$ for each nonnegative integer $k$.  Interestingly, in Theorem~\ref{THM:2}, we will see that for our problem, the optimal stopping policy is a threshold policy on $\Pi_k$ described by $8$ \emph{time-independent} thresholds.  This time-independence property greatly simplifies the computation.  The requirement on memory storage is also minimal, so this policy will be very simple to implement.  We comment that compared with the sequential testing problem, the cost function in our problem has a non-linear term $cT\Pi_T$.  So the proof for the sequential testing problem does not directly apply to our problem.  Nevertheless, we utilize an essential observation that the process $\{\Pi_k\}$ is a martingale with respect to $\{Z_k\}$ and still obtain a time-independent threshold policy.  The proof of Theorem~\ref{THM:2} is presented in Appendix~\ref{app:proof:thm2}.

\begin{theorem}
The optimal stopping time $T^*$ is
\begin{align}
T^*=\inf_{k>0}\left\{k: \Pi_{k}\notin\left(\pi_l^{(Z_k)},\pi_u^{(Z_k)}\right) \right\}.
\end{align} In other words, there exist positive values $\pi_l^{(z)},$ $\pi_u^{(z)},$ ($z\in\{0, 1, 2, 3\}$), independent of $T$, such that the algorithm declares the information to be news when $Z_k=z$ and $\Pi_k\leq \pi^{(z)}_l,$  and declares the information to be misinformation  when $Z_k=z$ and $\Pi_k\geq \pi^{(z)}_u.$
The thresholds $\pi_l^{(z)}$ and $\pi_u^{(z)}$ for $z=0, 1,2,3$ are determined by solving the following equations:
\begin{align}
\pi_l^{(z)}&=\sup_\pi\left\{\left.0\leq\pi\leq\frac{c_{\textnormal{I}}}{c_{\textnormal{I}}+c_{\textnormal{II}}}\right\vert s^{(z)}(\pi)=c_{\textnormal{II}}\pi\right\}\\
\pi_u^{(z)}&=\inf_\pi\left\{\left.\frac{c_{\textnormal{I}}}{c_{\textnormal{I}}+c_{\textnormal{II}}}\leq\pi\leq 1\right\vert s^{(z)}(\pi)=c_{\textnormal{I}}(1-\pi)\right\}
\end{align}
where $s$ is the solution of the Bellman equation below
\begin{align}
s^{(z)}(\pi)=\min\left\{g(\pi),E\left[\left.s^{(Z_{k+1})}(\Pi_{k+1})\right\vert \Pi_k=\pi,Z_k=z\right]+c\pi\right\}\label{eq:bellman}
\end{align}
and
$$g(\pi)=\min\{c_{\textnormal{II}}\pi,c_{\textnormal{I}}(1-\pi)\}.$$
\hfill{$\square$}\label{THM:2}
\end{theorem}

Note $E\left[\left.s^{(Z_{k+1})}(\Pi_{k+1})\right\vert \Pi_k=\pi,Z_k=z\right]$ in \eqref{eq:bellman} is understood as the expected cost to go starting from the next time step based on $s$ given the state in the current time step is $\pi$ and $z$.  In other words,
  \begin{align*}
    & \quad E\left[\left.s^{(Z_{k+1})}(\Pi_{k+1})\right\vert \Pi_k=\pi,Z_k=z\right]\\
    & = \sum_{z' = 0}^3s^{(z')}\left(\frac{\pi\alpha_1(z'\vert z)}{\pi\alpha_1(z'\vert z)+(1-\pi)\alpha_0(z'\vert z)}\right)\left(\pi\alpha_1(z'\vert z)+(1-\pi)\alpha_0(z'\vert z)\right).
  \end{align*}
So it does not depend on $k$.

%% file: alg-eval.tex
\section{\textsc{QuickStop}: The Quickest Misinformation Detection Algorithm}
\label{sec:alg}
From the results presented in the previous sections, we propose \textsc{QuickStop}, which includes the following components.

\begin{itemize}
\item {\bf  Training data:} Our algorithm needs labeled training data. The dataset should include a set of information spreading traces which are labeled as news or misinformation. Each user involved in the information trace has a feature vector. The information should also include the followee from whom a user retweeted the information.

\item {\bf Learning the information spreading model via the SVM classifier:} Given the labeled data, we first train an SVM classifier with the dataset that classifies information to news or misinformation. The input to the SVM classifier is the average feature vector of edges. Recall that the feature vector of edge $(v,u)$ is  {$({\bf V},{\bf U}).$} After training the SVM classifier, we use the classifier to classify the edges into four groups based on the edge feature vector. Note that SVM outputs an value between 0 to 1. In our experiments, we use the following mapping: $[0, 0.25]\Rightarrow 0,$ $(0.25, 0.5]\Rightarrow 1,$ $(0.5, 0.75]\Rightarrow 2,$ and $(0.75, 1]\Rightarrow 3.$ From the transition probabilities learned from the previous step, we calculate $\pi_l^{(z)}$ and $\pi_u^{(z)}$ according to Theorem \ref{THM:2}.

\item {\bf Quickest detection:} When monitoring information spreading, the algorithm updates $\Pi_k$ according to (\ref{eq:Pi}) when an event occurs, {where we set $\Pi_1=\pi_1$ which is the prior distribution of hypothesis $H_1$ according to the data.} The information is declared to be news when $\Pi_k\le\pi_l^{(Z_k)}$ and misinformation when $\Pi_k\ge\pi_u^{(Z_k)}$.
\end{itemize}

We remark that this algorithm combines a data-driven approach, which learns the underlying probabilistic model of information spreading in networks, and a model-driven approach, which identifies misinformation in a timely manner with the quickest detection formulation.

\textsc{QuickStop} consists of two parts: \textsc{QuickStop}-Training and \textsc{QuickStop}-Detection, whose pseudo-code can be found in Algorithms \ref{alg:quickstop-train} and \ref{alg:quickstop-detection}, respectively.

\begin{algorithm}[ht]
  \caption{\textsc{QuickStop}-Training (Offline)}\label{alg:quickstop-train}
  \begin{algorithmic}[1]
    \Require
        \Statex A set of information traces: $E=\{e_1,e_2,\cdots,e_n\}$ \Comment{$e_i$ is a sequence of users: {$\{u_t^{(e_i)}\},$}  where $t$ is the posting order of a user, $i$ is the index of the news trace}
        \Statex A set of labels: ${\bf l}=\{l_1,l_2,\cdots,l_n\}$ \Comment{$l_i\in\{0,1\}$ is the label of $e_i$ (0: news, 1: misinformation).}

    \State For the $t$th user who post the information $i$ (say user $u_t^{(e_i)}$),  obtain feature vector of the edge: {$({\bf V}_t^{(e_i)},{\bf U}_t^{(e_i)}).$}
    \State Compute $\tilde{\bf U}^{(e_i)}=\frac{1}{|e_i|-1}\left(\sum_{t=2}^{|e_i|}{{\bf V}_t^{(e_i)},}\sum_{t=2}^{|e_i|}{{\bf U}_t^{(e_i)}}\right)$ \Comment{$|e_i|$ is the cardinality of news trace $e_i$}
    \State Train edge classifier: $f(\cdot)$ using SVM with training dataset $(\tilde{\bf U},\bf{l})$
    \State Classify  edges in the traces, ${Z_t^{(e_i)}}\leftarrow {f({\bf V}_t^{(e_i)},{\bf U}_t^{(e_i)})}$
    \State Calculate the transition probabilities $$\alpha_j(z_1|z_2)=\frac{\sum_{i=1}^n\sum_{t=1}^{|e_i|-1}{\mathbb I}_{\{{Z^{(e_i)}_{t+1}=z_1,Z^{(e_i)}_{t}=z_2}\}}{\mathbb I}_{\{l_i=j\}}}{\sum_{i=1}^n\sum_{t=1}^{|e_i|-1}\mathbb I_{\{Z_t^{(e_i)} = z_2\}}{\mathbb I}_{\{l_i=j\}}},z_1,z_2\in\{0,1,2,3\}$$

     \State Initialize $\epsilon, \epsilon_0, m\leftarrow\frac{1}{
    \epsilon},\pi= \{\pi^1,\cdots,\pi^m\},c_{\text{I}},c_{\text{II}},c$ \Comment{$\epsilon$ and $\epsilon_0$ specify the quantization step size and the convergence tolerance}
    \For{$z=0,1,2,3$}
        \State $s_0^{(z)}(\pi^i)\leftarrow\min\{c_{\text{II}}\pi^i,c_{\text{I}}(1-\pi^i)\},i=1,\dots,m$
    \EndFor
    \For{$j=1,2,\dots$}    \Comment{Solve the Bellman equation using value iteration}
    \State $g(\pi^i)=\min\{c_{\text{II}}\pi^i,c_{\text{I}}(1-\pi^i)\},i=1,\dots,m$
    \For{$z=0,1,2,3$}
            \State $s_1^{(z)}(\pi^i)\leftarrow \min\left\{g(\pi^i),E\left[s_0^{(\tilde z)}(\tilde{\pi})\vert\pi^i,z\right]+c\pi^i\right\},i=1,\dots,m$ where \begin{align*}E\left[s_0^{(\tilde z)}(\tilde{\pi})\vert\pi^i,z\right]=&\sum_{k=0}^3s_0^{(k)}\left(\frac{\pi^i\alpha_1(k\vert z)}{\pi^i\alpha_1(k\vert z)+(1-\pi^i)\alpha_0(k\vert z)}\right)\\
            &\times(\pi^i\alpha_1(k\vert z)+(1-\pi^i)\alpha_0(k\vert z))
            \end{align*}
        \EndFor
        \If{$\|s_1(\pi)-s_0(\pi)\| \le \epsilon_0$}
            \State {\bf break}
        \Else
            \State $s_{0}(\pi)\leftarrow s_1(\pi)$
        \EndIf
    \EndFor
     \State $\pi_l^{(z)}\leftarrow\sup_{\pi^i}\left\{\left.0\leq\pi^i\leq\frac{c_{\text{I}}}{c_{\text{I}}+c_{\text{II}}}\right\vert s_0^{(z)}(\pi^i)=c_{\text{II}}\pi^i\right\},z\in\{0,1,2,3\}$
     \State $\pi_u^{(z)}\leftarrow\inf_{\pi^i}\left\{\left.\frac{c_{\text{I}}}{c_{\text{I}}+c_{\text{II}}}\le\pi^i\le 1\right\vert s_0^{(z)}(\pi^i)=c_{\text{I}}(1-\pi^i)\right\},z\in\{0,1,2,3\}$\Comment{Compute thresholds}
     \Ensure
        \Statex Edge classifier: $f(\cdot)$
        \Statex Transition probabilities: $\alpha_i(\cdot), \ i=0,1$
          \Statex Thresholds $\pi_l^{(z)} \hbox{and}\  \pi_u^{(z)},z\in\{0,1,2,3\}$
  \end{algorithmic}
\end{algorithm}

\begin{algorithm}[ht]
  \caption{\textsc{QuickStop}-Detection (Online)}\label{alg:quickstop-detection}
  \begin{algorithmic}[1]
    \Require
        \Statex Information trace: $y=\{y_1,y_2,\dots\}$ \Comment{$y_t$ is the $t$th   user in the information trace y }
        \Statex Edge classifier $f(\cdot)$
        \Statex Transition probabilities $\alpha_i(\cdot),i=0,1$
      \Statex Thresholds $\pi_l^{(z)} \hbox{and}\  \pi_u^{(z)},z\in\{0,1,2,3\}$
    \State Initialize $\Pi=\pi_1,k\leftarrow 2,Z_k^{(y)}=f({\bf V}_k^{(y)},{\bf U}_k^{(y)})$ \Comment{$\pi_1$ is the prior of $H_1$ (misinformation)}
    \While{$\Pi\in\left[\pi_l^{(Z_k^{(y)})},\pi_u^{(Z_k^{(y)})}\right]$}
      \State $k\leftarrow k+1$
       \State{For each user $y_k,$ obtain feature vector of edge: $({\bf V}_k^{(y)},{\bf U}_k^{(y)})$}
      \State $Z_k^{(y)}\leftarrow f({\bf V}_k^{(y)},{\bf U}_k^{(y)})$
      \State $\Pi\leftarrow \frac{\Pi \alpha_1(Z_k^{(y)}\vert Z^{(y)}_{k-1})}{(1-\Pi)\alpha_0(Z_k^{(y)}\vert Z^{(y)}_{k-1})+\Pi\alpha_1(Z^{(y)}_{k}\vert Z^{(y)}_{k-1})}$ \Comment{Compute $\Pi$}
      \State
    \EndWhile
    \State $T\leftarrow k$
    \If{$\Pi>\pi_u^{(Z_k^{(y)})}$}
      \State $\delta_T = 1$
    \ElsIf{$\Pi<\pi_l^{(Z_k^{(y)})}$}
      \State $\delta_T = 0$
    \EndIf
        \Ensure
        \Statex stopping time: $T,$ type of information: $\delta_T$
  \end{algorithmic}
\end{algorithm}

\subsection{Computational and Memory Complexities}

In the training part, we use an SVM classifier on $n$ information traces. In SVM, the feature space is obtained by using some mapping functions and the hyperplane is determined by a set of support vectors. Then the dimension of the feature space depends on the mapping function. The minimum computational complexity of training an SVM is $O(n^2),$ and may reach $O(n^3).$

The thresholds are calculated using the value iteration method. Let $\epsilon$ be the quantization step size of the state $\Pi_k$. During the value iteration, the terminal time depends on the quantization precision. The computational complexity for each iteration is $O(\frac{1}{\epsilon});$ the memory complexity is also $O(\frac{1}{\epsilon}).$ This step is done offline.

For the online misinformation detection part, the computational complexity per iteration and memory complexity are both $O(1).$ The algorithm  needs to store 8 threshold values and 32 transition probabilities. Each update of the state $\Pi_k$ only requires a few elementary operations.

\section{Performance Evaluation with Real-World Datasets}
We first evaluate the performance of \textsc{QuickStop} using the following real-world dataset.\label{sec:per_real}

\noindent {\bf The Weibo Dataset:} Sina Weibo is a Chinese microblogging website similar to Twitter.  The Weibo dataset we use is the one released in \cite{magaopra_16}, which includes 4,664 labeled information traces from Sina's community management center.\footnote{\url{https://service.account.weibo.com}} The dataset also includes user information such as the number of followees, the number of followers, the registration days, etc, which are used as user features in our algorithm. We remove information traces whose sizes are small. In particular, we keep the traces in which the information was retweeted by the followers of at least 50 distinct users. We further  balance the dataset by selecting 488 news traces and 488 misinformation traces. The average retweets per trace is 2,031, the largest trace includes 55,155 retweets, and the smallest one has 105 retweets. We used 80\% of the traces as training
data and the remaining as the testing data.

We compared \textsc{QuickStop} with the following misinformation detection algorithms aiming at early detection: (i) decision-tree-based methods \cite{casmenpob_11}; (ii) SVM-based methods with RBF kernel \cite{yanliuyu_12}; (iii) linear SVM-based models for time-series data \cite{magaozho_15}; (iv) Neural network-based methods with Recurrent Neural Networks (RNNs), or Convolutional Neural Networks (CNNs), or both for sequential data \cite{liuwu_18}; and (v) a comprehensive approach involving RNNs, Feedforward Neural Networks (FNNs), and singular value decomposition (SVD) for low-dimensional feature representation \cite{rucseoliu_17}. Note that {all these} methods are feature-based classification algorithms. {The first three algorithms \cite{casmenpob_11,yanliuyu_12,magaozho_15}} can take both user features and news content features as input. The algorithm proposed in \cite{liuwu_18} has three versions, RNN only, CNN only, and both. The algorithms use the sequential user features as the input to the neural networks. The algorithm in \cite{rucseoliu_17} uses an RNN to extract article features, an FNN to extract user features, and another FNN to integrate both user and article features for classification. \textsc{QuickStop}, on the other hand, only uses user features. In the evaluations, for the first three algorithms, we implemented two versions: one with only user features (i.e., the same set of user features used in \textsc{QuickStop}), and the other with both user and content features (so more features than \textsc{QuickStop}).  The ten different algorithms are summarized below.

\begin{itemize}

	\item {\bf DTC$_u$:} A Twitter information credibility method \cite{casmenpob_11} based on decision trees, with only user features.
	\item {\bf DTC$_a$:} A Twitter information credibility method \cite{casmenpob_11} based on decision trees, with both user and content features.

	\item {\bf SVM-RBF$_u$:} An SVM-based method with RBF kernel \cite{yanliuyu_12}, with only user features.
	
\item {\bf SVM-RBF$_a$:} An SVM-based method with RBF kernel \cite{yanliuyu_12}, with both user and content features.

	\item {\bf SVM-TS$_u$:} A linear SVM-based \cite{magaozho_15} method for time-series, with only user features.

	\item {\bf SVM-TS$_a$:} A linear SVM-based \cite{magaozho_15} method for time-series, with both user and content features.
	\item {\bf PPC\_R:} A variant of RNN \cite{liuwu_18} called Gated Recurrent Unit (GRU) for time-series data. The neural network has ~5,000 parameters.
    \item {\bf PPC\_C:} A CNN based method \cite{liuwu_18} for time-series data, which has ~800 parameters.
    \item {\bf PPC\_R+C:} A method in \cite{liuwu_18} that combines RNN and CNN, which has ~6,000 parameters.
	\item {\bf CSI:} A method proposed in  \cite{rucseoliu_17} that uses RNN for content feature extraction and FNN for user feature extraction. The three neural networks have 52,000 parameters in total.
\end{itemize}

We note that except \textsc{QuickStop},  all other algorithms mentioned require a pre-determined number of observations as input. \textsc{QuickStop} is an optimal stopping algorithm so it decides the number of observations needed in real time.

We remark that all the four neural network based methods (PPC\_R, PPC\_C, PPC\_R+C, and CSI) require a large number of samples for training. Therefore, we used 80\% of the entire 4,664 labeled traces for training the neural networks and then tested the performance on the same testing data as the other algorithms. The neural network based algorithms performed poorly when using the smaller training set as that in \textsc{QuickStop}.

\noindent{\bf Performance Metrics:} We considered the following performance metrics.
\begin{itemize}
\item Accuracy: the fraction of traces that are correctly identified.
\item False positive rate: the fraction of news classified as misinformation.
\item False negative rate: the fraction of misinformation classified as news.
\item Detection time of news: the average number of events required to declare news.
\item Detection time of misinformation: the average number of events required to declare misinformation.
\end{itemize}

\subsection{Numerical Results}

\begin{figure}[ht]
\centering
\includegraphics[scale = 0.7]{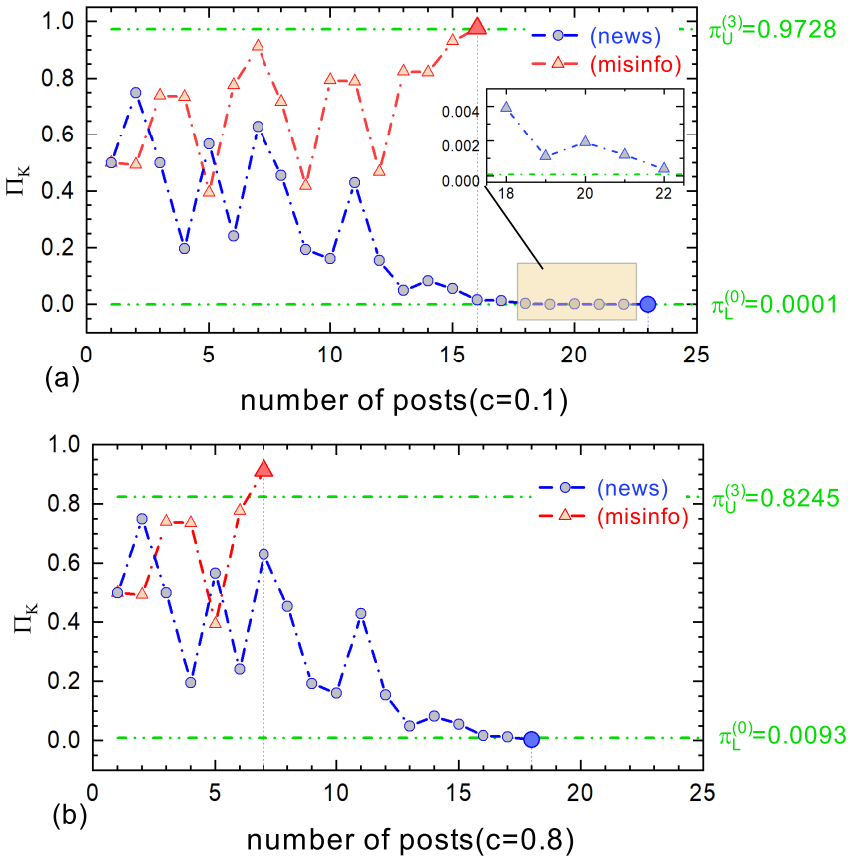}
\caption{Examples of $\Pi_k$ and stopping time $T$ under \textsc{QuickStop}}\label{fig:thresholds}
\end{figure}

\noindent{\bf Evolution of $\Pi_k$ under \textsc{QuickStop}:} Figure \ref{fig:thresholds} illustrates the evolution of $\Pi_k$ on two traces chosen from the Weibo dataset: one misinformation trace and one news trace.  We can see that the upper threshold becomes smaller and the lower threshold becomes larger when we increase the propagation cost from 0.1 to 0.8, and the algorithm stops earlier when $c=0.8$ than when $c=0.1$. Also it takes fewer number of observations to declare misinformation than news. With $c=0.8,$ it takes 7 observations to declare the misinformation and 18 observations to declare the news. Similar trends can be observed on most of the traces.

\begin{figure}[ht]
\centering
\includegraphics[scale = 1]{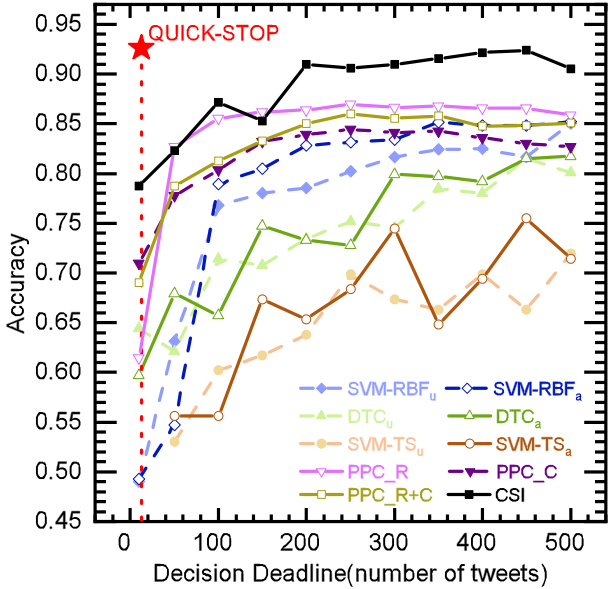}
\caption{Performance of Early Misinformation Detection under Different Decision Deadlines (based on the Weibo Data)}\label{fig:result-early}
\end{figure}

\begin{figure*}[ht]
\centering
\includegraphics[scale = 0.7]{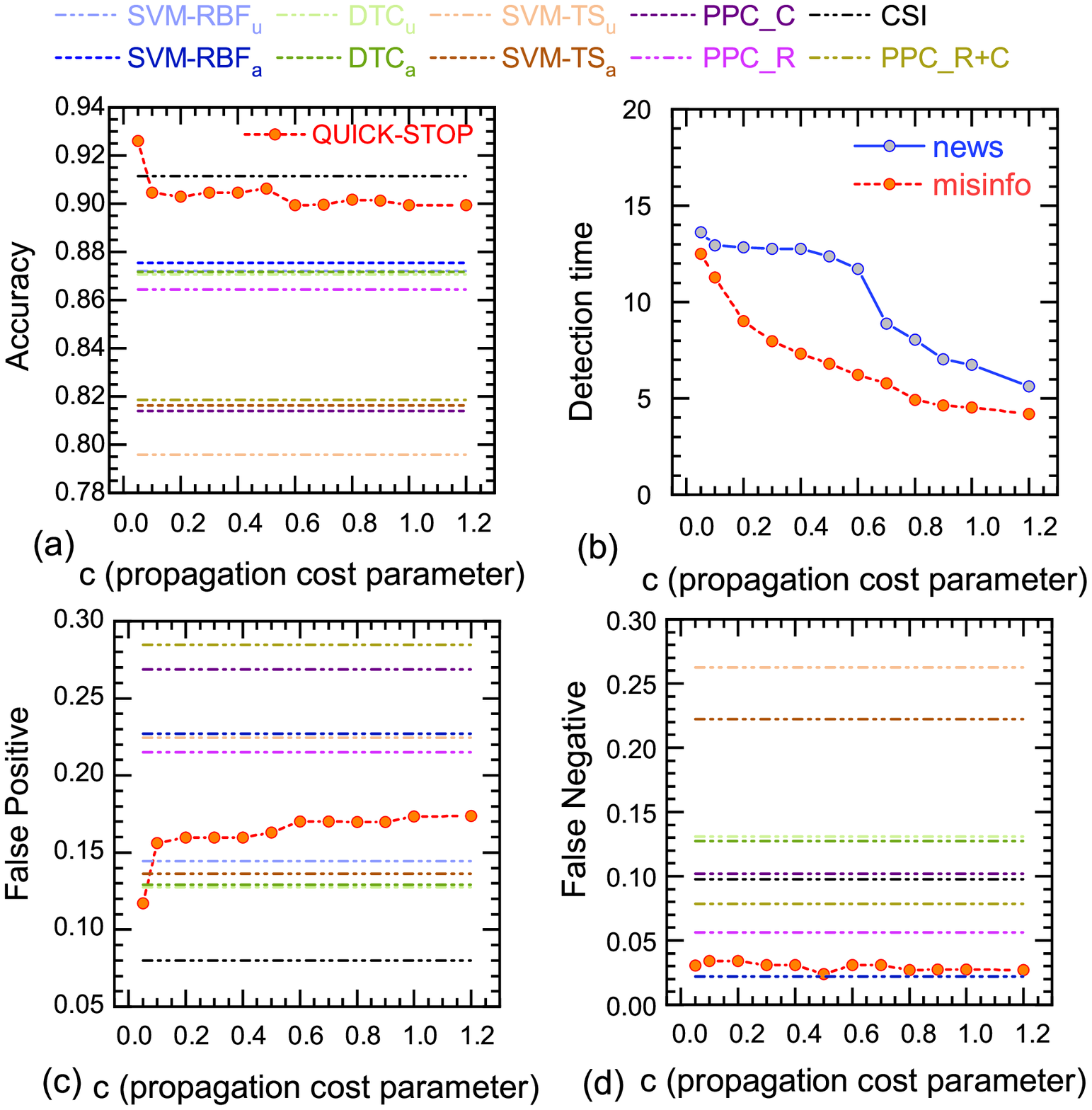}
\caption{Performance of \textsc{QuickStop} under Different Choices of Parameter $c$ (based on the Weibo Data)}\label{fig:result-weibo}
\end{figure*}

Figures  \ref{fig:result-early} and \ref{fig:result-weibo} summarize the performances of \textsc{QuickStop} and the other ten algorithms. In Figure \ref{fig:result-early}, \textsc{QuickStop} uses parameters $c_{\text{I}}=c_{\text{II}}=10$ and $c=0.05;$ and the $x$-axis is the number of tweets used by the other ten algorithms, varying from 10 to 500.  Note that when the number of observations in a trace is less than the decision deadline, then the full trace was used as the input. In Figure \ref{fig:result-weibo}, we varied the parameter $c$ of \textsc{QuickStop} from 0.05 to 1.2 with step size 0.05. In Figure \ref{fig:result-weibo}, all ten other algorithms used full Weibo traces as input. The key observations are summarized below.
\begin{itemize}
	\item {\bf High Accuracy:} Figure~\ref{fig:result-weibo} (a) shows that the accuracy of \textsc{QuickStop} only with user features is substantially higher than other algorithms even when other algorithms use both user features and content features.  Specifically, \textsc{QuickStop} with $c=0.05$ achieves higher accuracy than other algorithms with 500 observations with less than $15$ observations on average. Under \textsc{QuickStop}, as $c$ increases, the accuracy decreases but the number of observations used decreases as well, which is the trade-off between accuracy and speed.

\item {\bf Quick Detection:} Quickest misinformation detection is the key objective of our algorithm.  Figure \ref{fig:result-early} shows that the accuracy of \textsc{QuickStop} in comparison with the other algorithms. \textsc{QuickStop} with $c=0.05$ achieves an accuracy of 0.93 with $15$ observations on average while the accuracies of all other algorithms are lower than {0.93} even with 500 observations. Note that {four} of the {ten} algorithms include content features which are not used in \textsc{QuickStop}. 

\item {\bf Low False Negative:} In almost all cases, the false negative rate of \textsc{QuickStop} is lower than the false positive rate. This is because with the discriminative propagation cost, \textsc{QuickStop} is more aggressive on declaring misinformation than news in order to minimize the propagation cost.  We also remark that CSI, which involves 52,000 parameters, has an accuracy close to \textsc{QuickStop} when using entire traces, but its false negative rate is much higher than \textsc{QuickStop} (0.097 versus 0.031).
\end{itemize}

{The experimental results show that \textsc{QuickStop} detects misinformation faster and more accurately than other algorithms. We believe it is because \textsc{QuickStop} specifically models and utilizes the Markovian structure of the problem, and is based on the optimal stopping rule. The other algorithms were not optimized for the stopping time, nor do they have theoretical guarantees.}

\section{Evaluation with Synthetic Data}
We further evaluate the algorithm with synthetic network and information spreading data. We construct a network with $500$ nodes using the preferential attachment model \cite{simher_55}. Our network includes two types of nodes: gossipers and messengers, where gossipers are more likely to spread misinformation than messengers. When a new node joins the network, it is assigned a type uniform at random, and then connects to three existing nodes in the network, i.e. forming three edges. For each edge, the new node first decides whether to connect to a node of the same type (with probability 0.7) or a node of different type (with probability 0.3). After deciding the type, say it chooses to connected to a gossiper, the new node selects a gossiper among all existing gossipers with probability proportional to their degrees. We define the edge types as follows: 0 - (messenger, messenger), 1 - (gossiper, messenger), 2 - (messenger, gossiper) and 3 - (gossiper, gossiper). We simulated the information spreading using the continuous-time SI model. For each set of parameters, we create $500$ traces. Each trace was flagged as news with $\pi_0=\pi_1=0.5$. The probabilities that an article is retweet over a given edge under the SI model are summarized in Table \ref{tab:edge-infospreading}. From example, news spreads from a messenger to another messenger with probability 0.9, spreads from a gossiper to messenger with probability 0.7, misinformation spreads from a messenger to another messenger with probability 0.1, and from a gossiper to another gossiper with probability 0.9. 

\begin{table}[ht]
	\centering
	\begin{tabular}{|c|c|c|c|c|}
		\hline
		&0&1&2&3\\
		\hline
		News& 0.9& 	0.7& 0.3 & 	0.1 \\
		\hline
		Misinformation& 0.1& 	0.2& 	0.7& 	0.9\\
		\hline
	\end{tabular}
	\caption{Probability of Information Spreading over Different Edge Types}\label{tab:edge-infospreading}
\end{table}

The objective of this evaluation with the synthetic data is to evaluate the robustness of the online \textsc{QuickStop}-Detection with classification errors. With the synthetic data, the edge types are known so we can control the edge classification errors by random flipping the edge types and evaluate the performance of \textsc{QuickStop}-Detection with respect to classification errors.

\begin{figure}[ht]
\centering
\includegraphics[scale = 0.7]{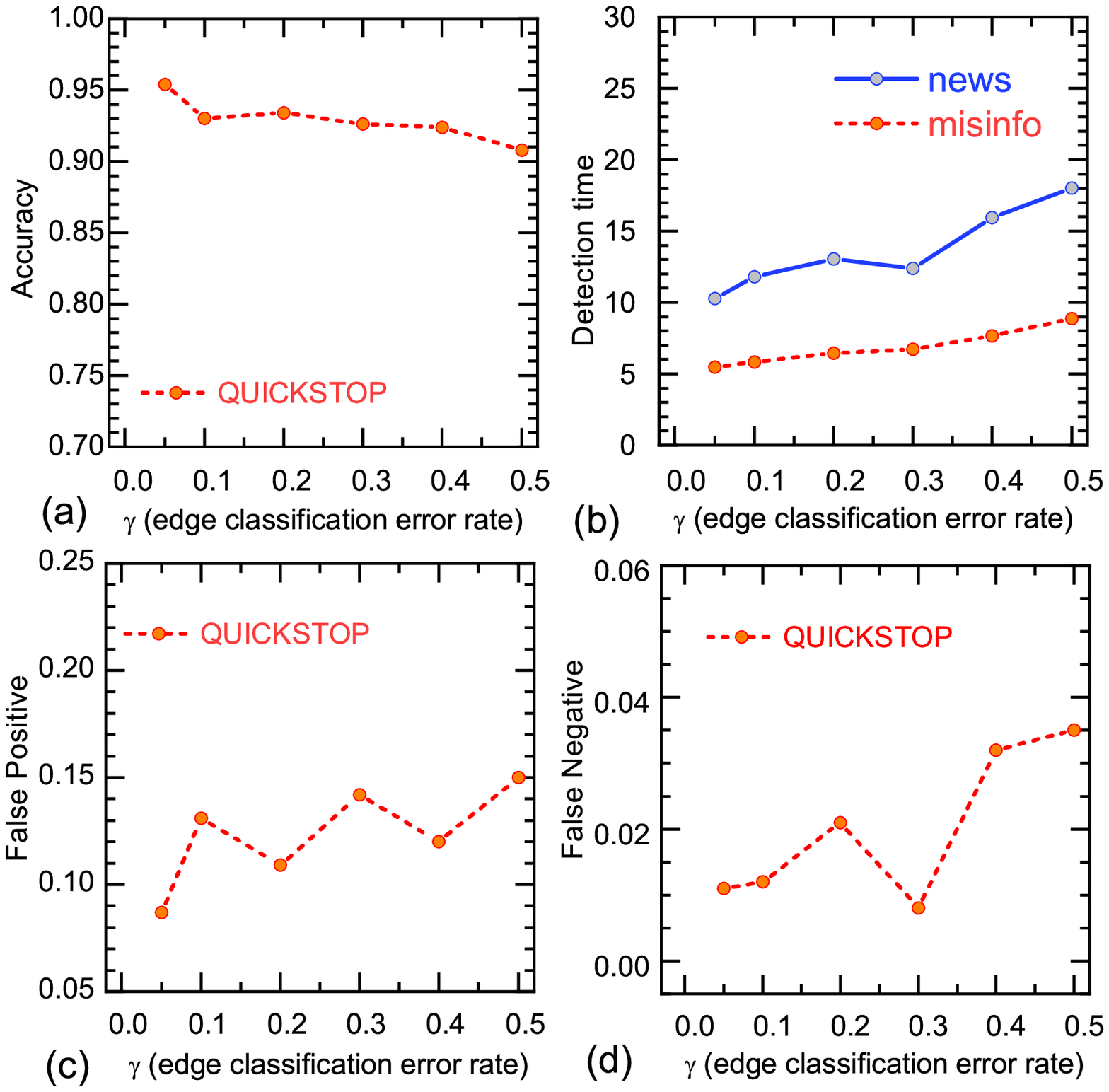}
\caption{Performance of \textsc{QuickStop} with the synthetic Data with classification errors}\label{fig:simulation_noise_result}
\end{figure}

Figure \ref{fig:simulation_noise_result} shows the performance of \textsc{QuickStop} with different classifcation errors. We introduced edge classification errors such that the type of an edge is correctly classified with probability $\gamma$ and misclassified with probability $1-\gamma.$ We varied $\gamma$ from $0.05$ to $0.5$.  In Figure~\ref{fig:simulation_noise_result}, we used $c_{\text{I}}=c_{\text{II}}=10$ and $c=0.3$ for \textsc{QuickStop}.
\begin{itemize}
\item {\bf Robust to Learning Errors:}  We can observe that even when 50\% edges are not correctly classified, \textsc{QuickStop} still has an accuracy close to 91\%, which demonstrates the robustness of the detection to modeling errors.
\end{itemize}

%% file: related.tex
\section{Related Work}
\label{sec:related}
As we pointed out at the beginning of the introduction, government, industry and academia have made great efforts to combat misinformation. This section focuses on new developments on misinformation detection with machine-learning and data-mining methods in the research community.

We have discussed several early detection algorithms and compared their performance with \textsc{QuickStop}. We now focus on other related work. The algorithm developed in \cite{qazrosrad_11} detects whether a post is similar to one of the posts (topics) that are known to be misinformation; and declares it as misinformation if so.  A line of work \cite{taksofint_12,hamdia_16,hamdia_15} analyzes similar models and knowledge/content-based  detection algorithms.  These approaches are effective for detecting whether a post is associated with misinformation already identified, but not suitable for detecting new misinformation.  \citep{bancaesod_07,magwan_10,ciamshiroc_15,shiwen_16} exploit open fact-checking sources (such as DPpedia, Wikipedia, etc) to validate the truthfulness of news articles.   Viewpoints of users towards news articles such as ``like'' and ``dislike'' have also been used in the literature to infer the veracity of a news article. For example, \citep{tacbaldel_17} classifies Facebook posts as hoax or non-hoaxes based on the set of users who ``liked'' them. The work \cite{jincaoyon_16} uses a topic model to discover viewpoint values from tweets and evaluated the credibility of relevant posts based on these viewpoints. 

In \cite{casmenpob_11}, a comprehensive data-mining approach has been proposed for determining the veracity of social media contents. They considered four categories of features: message-based, user-based, topic-based, and propagation-based features to study information credibility, and proposed a PageRank-like credibility analysis method to verify the credibility of twitter events. The features used in \cite{casmenpob_11} have later been used in other papers \citep{yanliuyu_12,liunouli_15,gupzhahan_12}.  In \cite{kwochajun_17}, the authors argued that features vary over time. They reported that linguistic features are effective for detecting rumor even at the early stage of information spreading. A model for time-varying features has been proposed in \cite{magaozho_15}. \cite{wuyanzhu_15} explores the use of the features of the message propagation trees for detecting misinformation. \cite{chuban_16} analyzes six categories of features: comprehensibility, sentiment, time-orientation, quantitative details, writing style, and topic. \cite{derbonlia_17} analyzes users' stance in their tweets to evaluate the credibility of information. \cite{chazhansza_16} studies the characteristics of users who often post misinformation, and proposes that after identifying these users, a news article is likely to be misinformation if it spreads among these users. \cite{vos_15} proposes a misinformation detection algorithm with dynamic time wrapping and hidden Markov models based on three categories of features (linguistic, user identities and temporal propagation related features).  Recently, deep neural network (CNN, RNN, and FNN) based methods have also been used for misinformation detection \cite{rucseoliu_17,liuwu_18}.

Users play the central role in information diffusion in social networks. Their social engagements such as sharing, forwarding, commenting are considered to be auxiliary information for improving fake news detection. \cite{tscsebadi_18} uses users' flags of fake news as signals and leverages community for misinformation detection by learning the users' flagging accuracy. Online social network users who intentionally spread misinformation can be divided into three categories: (1) bots, software apps that run automated scripts\footnote{\url{https://en.wikipedia.org/wiki/Internet_bot}} (2) trolls, persons who like to provoke others, and (3) cyborgs\footnote{\url{https://en.wikipedia.org/wiki/Internet_troll}}, accounts registered to run automated programs that mimic human behaviors \cite{shusliwan_17}. \cite{chexuehon_18,shaciavar_17} analyze the behavior patterns of bots and trolls in misinformation propagation. In \cite{chugiawan_12}, an automated method is proposed for classifying the users into the three categories mentioned above. In \cite{morwunaz_16}, bot detection is studied. \cite{shuwanliu_18} analyzes the users' role in spreading information and concludes  that (1) some specific users are more likely to believe in misinformation than real news; (2) these users have different features form other users. These two key observations motivated the edge-based model considered in this paper. \cite{abbliu_13} proposes a method for measuring user credibility in information spreading for misinformation detection. The spread of rumors and misinformation has also been studied in \cite{vosroyara_18,jindousar_13,friadaeck_14}, where it has been shown that misinformation and news have different spreading patterns and structures. In this paper, we consider both edge profiles (the edge classification) and spreading patterns (the Markovian spreading model) in \textsc{QuickStop} to design a highly efficient misinformation detection algorithm. Different from existing work, \textsc{QuickStop} is an optimal stopping algorithm that optimizes the number of observations in realtime and makes the quickest decision on misinformation detection. {Finally, recent algorithms for distinguishing epidemics from random infection (e.g., \cite{milcarman_13}) and for locating information sources (e.g., \cite{shazam_11})  can also help detect misinformation. A comprehensive review of diffusion source localization can be found in \cite{ZhuYin_18}.}

%% file: conclusions.tex
\section{Conclusions}
In this paper, we proposed a quickest misinformation detection algorithm, named \textsc{QuickStop}. We formulated the problem as an optimal stopping problem with a asymmetric cost function towards misinformation. We proved that the problem is a Markov optimal stopping problem and showed that the solution is a threshold-based stopping rule based on the martingale theory.   Our numerical results with a  real-world data demonstrated that \textsc{QuickStop} outperforms existing algorithms even though the latter use 10 times (sometimes 50 times) more observations and use more features. Our numerical evaluation with the synthetic data showed that the algorithm is robust to edge classification errors.

%% file: app.tex
\appendix
\section{Appendices}
\subsection{Proof of Theorem \ref{THM:1}}
\label{app:proof:thm1}
We first show that $E[cT\mathbb{I}_{H_1}]=E\left[cT\Pi_{T}\right]$ when $T$ is a stopping time.
\begin{align*}
E[cT\mathbb{I}_{H_1}]&=E[E[cT\mathbb{I}_{H_1}\vert T]]\\
&=\sum_{k=1}^\infty ckE\left[\mathbb{I}_{H_1}\vert T=k\right]\Pr(T=k).
\end{align*}
Since $T$ is a stopping time based on $Z_1, \cdots, Z_T,$ we further have
\begin{align*}
E\left[\mathbb{I}_{H_1}\vert T=k\right]&=E\left[E\left[\mathbb{I}_{H_1}\vert Z_1,\ldots,Z_k\right]\vert T=k\right]\\
&=E\left[\Pi_k\vert T=k\right].
\end{align*} Therefore, we have
\begin{align*}
E[cT\mathbb{I}_{H_1}]&=\sum_{k=1}^\infty ckE\left[\Pi_k\vert T=k\right]\Pr(T=k)=E\left[cT\Pi_T\right].
\end{align*}

For any $T\in\mathcal{T}$, it is well known (see for example \cite{poovinhad_09}) that
$$\inf_{\delta_T}c_e(\delta_T)=E\left[\min\{c_{\text{II}}\Pi_{T},c_{\text{I}}(1-\Pi_{T})\}\right].$$ We next present the proof tailored for our problem for the completeness of the paper.

Note that the equation is obvious when $\pi_1=0$ or $\pi_1=1$, so we only consider the case $\pi_1\in(0,1)$.  Recall that
\begin{align*}
&c_e(\delta_T)\\
=&(1-\pi_1)c_{\text{I}}\Pr(\delta_T=1|H_0)+\pi_1 c_{\text{II}}\Pr(\delta_T=0|H_1)\\
=&c_{\text{I}}\Pr(\delta_T=1,H_0)+c_{\text{II}}\Pr(\delta_T=0,H_1)\\
=&\sum_{k=1}^\infty \left(c_{\text{I}}\Pr(\delta_T=1, H_0\vert T=k)+c_{\text{II}}\Pr(\delta_T=0, H_1\vert T=k)\right)\Pr\left(T=k\right)\\
=&\sum_{k=1}^\infty \left(c_{\text{I}}\Pr(\delta_k=1, H_0\vert T=k)+c_{\text{II}}\Pr(\delta_k=0, H_1\vert T=k)\right)\Pr\left(T=k\right)\\
=&\sum_{k=1}^\infty \left(c_{\text{I}}E\left[E\left[\mathbb{I}_{\delta_k=1}\mathbb{I}_{H_0}\vert Z_1, \cdots, Z_k\right]\vert T=k\right]+\right.\\
&\quad \left.c_{\text{II}}E\left[E\left[\mathbb{I}_{\delta_k=0}\mathbb{I}_{H_1}\vert Z_1, \cdots, Z_k\right]\vert T=k\right]\right)\Pr\left(T=k\right)\\
=&\sum_{k=1}^\infty E\left[ \left.\left(c_{\text{I}}\mathbb{I}_{\delta_k(Z_1, \cdots, Z_k)=1}(1-\Pi_k)+c_{\text{II}} \mathbb{I}_{\delta_k(Z_1, \cdots, Z_k)=0}\Pi_k\right) \right| T=k\right]\\
&\times\Pr\left(T=k\right)\\
&\geq_{(a)} \sum_{k=1}^\infty E\left[ \left.\min\left\{c_{\text{I}}(1-\Pi_k), c_{\text{II}} \Pi_k\right\} \right| T=k\right]\Pr\left(T=k\right)\\
&=E\left[\min\left\{c_{\text{I}}(1-\Pi_T), c_{\text{II}} \Pi_T\right\}\right],
\end{align*} where the inequality $(a)$ becomes equality when the algorithm declares $H_1$ when $c_{\text{I}}(1-\Pi_T) \leq c_{\text{II}}\Pi_T$ and declares $H_0$ otherwise.

\subsection{Proof of Theorem \ref{THM:2}}
\label{app:proof:thm2}
We define the following value function for $n\ge 1$
$$s_n(\pi,z)=\inf_{T\in\mathcal{T},T\geq n}E\left[g(\Pi_T)+cT\Pi_T\vert \Pi_n=\pi,Z_n=z\right].$$
Then $s_n(\pi, z)$ is the minimum expected total cost if one is only allowed to stop at or after time step $n$ given the state at $n$.  Note $\mathcal T = \{T\in\mathcal T\colon T\ge 1\}$.  Then the minimum expected total cost over the prior $\pi_0$ is
$$s_1^*\triangleq\inf_{T\in\mathcal{T}}E\left[g(\Pi_T)+cT\Pi_T\right]=\frac{1}{4}\sum_{z=0}^3s_1(\pi_0,z),$$ where we use the fact that $\Pi_1=\Pi_0=\pi_0$ as the first observation $Z_1$ does not provide any information about the type of the information.

Now according to the optimality principle of dynamic programming,
$$s_k(\pi,z)=\min\left\{g(\pi)+ck\pi,E\left[\left.s_{k+1}(\Pi_{k+1}, Z_{k+1})\right\vert \Pi_k=\pi,Z_k=z\right]\right\},$$ where $\{\Pi_k\}$ is a random process defined by $\{Z_k\}$ as in equation~\eqref{eq:Pi}.

We next show that $\{\Pi_{k}\}$ is a martingale with respect to $\{Z_k\}$. Define $\mathcal{F}_k=\sigma(Z_1, \cdots, Z_k)$, which is the $\sigma$-algebra generated by $Z_1$, \dots, $Z_k$.  We have
\begin{align*}
 E\left[\Pi_{k+1}\vert \mathcal{F}_k\right] =\sum_{z = 0}^3E\left[\Pi_{k+1}\vert\mathcal{F}_k,Z_{k+1}=z\right]\Pr(Z_{k+1}=z\vert{\mathcal F}_k). \end{align*}
Since
\begin{align*}
&\Pr(Z_{k+1}=z\vert\mathcal{F}_k)\\
=&\Pr(Z_{k+1}=z\vert\mathcal{F}_k,H_1)\Pr(H_1\vert \mathcal{F}_k)+\Pr(Z_{k+1}=z\vert\mathcal{F}_k,H_0)\Pr(H_0\vert \mathcal{F}_k)\\
=&\alpha_1(Z_{k+1}=z\vert Z_k)\Pi_k+\alpha_0(Z_{k+1}=z\vert Z_k)(1-\Pi_k),
\end{align*}
we have
\begin{align*}
&E\left[\Pi_{k+1}\vert \mathcal{F}_k\right] \\
=&\sum_z \frac{\Pi_k\alpha_1(Z_{k+1}=z\vert Z_k)}{\Pi_k\alpha_1(Z_{k+1}=z\vert Z_k)+(1-\Pi_k)\alpha_0(Z_{k+1}=z\vert Z_k)}\\
&\times\left(\alpha_1(Z_{k+1}=z\vert Z_k)\Pi_k+\alpha_0(Z_{k+1}=z\vert Z_k)(1-\Pi_k)\right)\\
=&\sum_z \Pi_k\alpha_1(Z_{k+1}=z\vert Z_k)\\
=&\Pi_k.
\end{align*}
For $n\ge 1$ let $\Pi_k' = \Pi_{k+n-1}$ and $Z_k' = Z_{k+n-1}$ for all $k \ge 1$.  Then
\begin{align*}
  & \quad s_n(\pi, z)\\
  & =\inf_{\substack{T\in\mathcal{T}\\T-n+1\geq 1}}E\left[g(\Pi_T)+c(T-n+1)\Pi_T+c(n-1)\Pi_T\vert \Pi_n=\pi,Z_n=z\right]\\
  & = \inf_{\substack{T'\in\mathcal T\\T'\ge 1}}E[g(\Pi_{T'}')+cT'\Pi_{T'}'\vert \Pi_1' = \pi, Z_1' = z]+c(n-1)\pi\\
  & = s_1(\pi, z)+c(n-1)\pi.
\end{align*}
In other words, because the posterior probability $\{\Pi_k\}$ is a martingale with respect to the observations $\{Z_k\}$, every time step passed before time $n$ (when one is allowed to stop and make a decision) incurs a constant additive cost of $c\pi$ to the minimum expected total cost.

Now define
\begin{equation}
  \label{eq:valueFunction}
  s^{(z)}(\pi) = s_1(\pi, z)-c\pi.
\end{equation}
Then for any $k\ge 1$,
\begin{align*}
&\quad s^{(z)}(\pi)\\
&=s_k(\pi, z)-ck\pi\\
&=\min\left\{g(\pi),E\left[\left.s^{(Z_{k+1})}({\Pi}_{k+1})+c(k+1){\Pi}_{k+1}\right\vert\Pi_k=\pi,Z_k=z\right]-ck\pi\right\}\\
&=\min\left\{g(\pi),E\left[\left.s^{(Z_{k+1})}({\Pi}_{k+1})\right\vert\Pi_k=\pi, Z_k=z\right]+c\pi\right\}.
\end{align*}
Hence $s$ as defined in \eqref{eq:valueFunction} satisfies the Bellman equation \eqref{eq:bellman}.

Note that $g(\pi)+ck\pi$ is the cost when the information type is declared at iteration $k$ given $\Pi_k=\pi,$ and $E\left[\left.s^{(z)}({\Pi}_{k+1})\right\vert\Pi_k=\pi, Z_k=z\right]+c(k+1)\pi$ is the minimum cost the information type is declared after iteration $k$ given $\Pi_k=\pi.$ Therefore, at optimal stopping time $T,$ we have
$$s^{(z)}(\pi)+ck\pi=g(\pi)+ck\pi$$ i.e., $$s^{(z)}(\pi)=g(\pi).$$ Furthermore, if $s^{(z)}(\pi)=g(\pi)$ and $c_{\text{II}}\pi<c_{\text{I}}(1-\pi),$ then $s^{(z)}(\pi)=c_{\text{II}}\pi,$ so the information is declared to be news; otherwise, it is declared to be misinformation. Therefore, after solving $s^{(z)})(\pi),$ we have $$\pi^{(z)}_l=\sup_\pi\left\{\pi: s^{(z)}(\pi)=c_{\text{II}}\pi\right\},$$ and
$$\pi^{(z)}_u=\inf_\pi \left\{\pi: s^{(z)}(\pi)=c_{\text{I}}(1-\pi)\right\}.$$